\documentclass[epj]{svjour}

\usepackage{graphicx}
\usepackage{fancyhdr}
\def\makeheadbox{{
 }}
\def\mytitle{My title}
\def\myauthors{My name}
\def\mytype{My type of session}
\def\mysession{My session}

\def\mytitle{Radiative Penguin Decays at the $B$ Factories} 
\def\myauthors{Karsten K\"oneke}    
\def\mytype{Contributed Talk}
\def\mysession{Flavor Physics}

\def\ifmath#1{\relax\ifmmode#1\else$#1$\fi}

\def\eg{{\it e.g.}}

\def\wrt{{\it w.r.t.}}

\def\babar{\mbox{\sl B\hspace{-0.4em} {\scriptsize\sl A}\hspace{-0.4em} B\hspace{-0.4em} {\scriptsize\sl A\hspace{-0.1em}R}}}
\def\belle{\mbox{Belle}}

\def\mev  {\ifmath{\mbox{\,Me\kern -0.08em V}}} 
\def\gev  {\ifmath{\mbox{\,Ge\kern -0.08em V}}} 
\def\gevc {\ifmath{\mbox{\,Ge\kern -0.08em V$\!/c$}}} 
\def\mevc {\ifmath{\mbox{\,Me\kern -0.08em V$\!/c$}}} 
\def\gevcc{\ifmath{\mbox{\,Ge\kern -0.08em V$\!/c^2$}}} 
\def\mevcc{\ifmath{\mbox{\,Me\kern -0.08em V$\!/c^2$}}} 

\def\mes   {\ifmath{m_{ES}}}
\def\de    {\ifmath{\Delta E}}
\def\vtdvts {\ifmath{\left| V_{td}/V_{ts}\right|}}

\def\to{\rightarrow}

\def\bd   {\ifmath{B^0_d}}
\def\bs   {\ifmath{B^0_s}}
\def\bsss {\ifmath{B^{0\left(*\right)}_s}}
\def\bss  {\ifmath{B^{0*}_s}}
\def\brog {\ifmath{B \to \left( \rho / \omega \right) \gamma}}
\def\brg   {\ifmath{B \to \rho\gamma}}
\def\brzg  {\ifmath{B^0 \to \rho^0\gamma}}
\def\brpg  {\ifmath{B^+ \to \rho^+\gamma}}
\def\bog   {\ifmath{B^0 \to \omega\gamma}}
\def\bsg  {\ifmath{b\to s\gamma}}
\def\bkg   {\ifmath{B \to K^{*}\gamma}}
\def\bkzg  {\ifmath{B^0 \to K^{*0}\gamma}}
\def\bsphig  {\ifmath{\bs \to \phi\gamma}}
\def\bsgg    {\ifmath{\bs \to \gamma\gamma}}
\def\bdpi    {\ifmath{B \to D \pi}}

\def\bsgg    {\ifmath{\bs \to \gamma \gamma}}

\def\BFrp{1.10^{+0.37}_{-0.33}\pm 0.09} 
\def\BFrz{0.79^{+0.22}_{-0.20}\pm 0.06} 
\def\BFom{0.40^{+0.24}_{-0.20}\pm 0.05} 
\def\BFav{1.25^{+0.25}_{-0.24}\pm 0.09} 
\def\BFavrhorho{1.36^{+0.29}_{-0.27}\pm0.10} 
\def\VtdVtsvalEx{0.194 ^{+0.015} _{-0.014} {\rm (exp.)} \pm 0.014 {\rm (th.)} } 
\def\VtdVtsBabarRhopzEx{0.201 \pm 0.016 {\rm (exp.)} \pm 0.015 {\rm (th.)} } %
\def\VtdVtsCDFEx{0.208 ^{+0.001} _{-0.002}{\rm (exp.)} ^{+0.008} _{-0.006}{\rm (th.)} } 

\def\epem  {\ifmath{e^+e^-}}
\def\BB    {\ifmath{B\overline{B}{}}} 
\def\FourS {\ifmath{\Upsilon{\rm( 4S)}}} 
\def\FiveS {\ifmath{\Upsilon{\rm( 5S)}}} 
\def\FourSBB {\ifmath{\Upsilon (4S)\to B \overline{B}}}
\def\eebb  {\ifmath{\epem \to \FourS \to \BB}}
\def\eeff     {\ifmath{\epem \to f \overline{f}}}

\begin{document}
\title{Radiative Penguin Decays at the $B$ Factories}
\author{Karsten K\"oneke\inst{1}\thanks{\emph{Email:} kkoeneke@alum.mit.edu} (for the \babar\ and Belle
Collaborations)
}                     
%
%
\institute{Massachusetts Institute of Technology, Laboratory for
Nuclear Science, Cambridge, Massachusetts 02139, USA}
%
\date{}
\abstract{ In this article, I review the most recent results in
radiative penguin decays from the $B$ factories Belle and \babar.
Most notably, I will talk about the recent new observations in the
decays \brog, a new analysis technique in \bsg, and first
measurements of radiative penguin decays in the \bs\ meson system.
Finally, I will summarize the current status and future prospects of
radiative penguin $B$ physics at the $B$ factories.
} 
\maketitle
\section{Introduction}
\label{intro}

Now, at the verge of the Large Hadron Collider (LHC) becoming
operational, an exciting new era of discovery in high-energy
particle physics is anticipated. There is a whole series of new
physics models attempting to fix some of the shortcomings of the
Standard Model (SM) of elementary particle physics, most notably the
hierarchy problem of the Higgs sector. Low energy supersymmetry
(SUSY) is probably the most popular extension of the SM. But all
these new models introduce new free parameters. Current precision
experiments can be used to constrain the allowed space of these
additional new parameters and therefor guide the searches at the
LHC.

One very interesting class of decays are so-called penguin decays
where the leading order SM decay contribution is described by a
one-loop Feynman diagram. Here, I will concentrate on penguin decays
of $B$ mesons where a photon is radiated off.

Unknown particles that do not exist in the SM can contribute
considerably to these decays, \eg\ by appearing in this loop, and
thus shift measurable observables. But if the measurements agree
with SM based theoretical predictions, constraints on various beyond
the SM physics scenarios can be extracted.

There is a large interest in $B$ meson decays since they provide an
excellent window into the physics of quark mixing. In the SM, this
is parameterized by four numbers that appear as the four independent
parameters of the Cabibbo-Kobayashi-Maskawa (CKM) quark mixing
matrix. A stringent test of the SM is performed by overconstraining
these four parameters with independent measurements, of which many
can be done by studying $B$ meson decays.

Hundreds of millions of $B$ mesons have been produced in the process
\eebb\ at the two energy-asymmetric \epem\ colliders KEKB and
PEP-II. The measurements reviewed in this paper are performed with
the \belle\ and \babar\ detectors \cite{Belle:2000cg,Aubert:2001tu}
situated at these two $B$ factories. In Sec.~\ref{sec::recon}, I
review common reconstruction techniques. In
Sec.~\ref{sec::rhogamma}, I will talk about the decays \brog. I
introduce a new experimental approach to measuring the decay
\bsg\footnote{Charge conjugation is implied throughout this
article.} in Sec.~\ref{sec::bsgamma}. In Sec.~\ref{sec::fives}, I
report on results of radiative penguin \bs\ decays using a data
sample collected at the \FiveS\ resonance. Finally, I conclude in
Sec.~\ref{sec::conclusions}.

\section{Common reconstruction techniques}
\label{sec::recon}

$B$ mesons\footnote{All particles mentioned in the experimental part
of this paper are actually ``candidates'' in the sense that their
true identity is only assumed.} are reconstructed in two standard
variables, \mes\ and \de. The first is defined as
$\mes = \sqrt{\frac{s}{4} - \left|\vec{p}^{*}_{B} \right|^{2}}$,
where $s$ is the center-of-momentum (CM) energy of the \epem\
collision and $\vec{p}^{*}_{B}$ is the measured three-vector of the
reconstructed $B$ meson, also in the CM frame. The second important
variable is defined as
$\de =  E_{B}^{*} - \sqrt{s}/2$, where $E_{B}^{*}$ is the measured
CM energy of the reconstructed $B$ meson.

Due to the small mass difference between $m_{\FourS} - 2 m_{B}$, the
two $B$ mesons are almost at rest in the CM frame. They decay thus
rather spherically symmetric in that frame. This is different for
the so-called light-quark continuum background (\eeff, where $f = u,
d, s, c, \tau$) where the light quarks (or leptons) have a large
momentum in the CM frame. Their hadronization and decay is thus
relatively collimated along the decay axis. These event shape
differences are used in various ways in all analysis described
below.

\section{The Unitarity Triangle and \brog}
\label{sec::rhogamma}

Besides the generic interests in radiative penguin decays mentioned
in the introduction, this decay is also interesting due to its
implications to measuring CKM matrix elements. If the experimental
branching fraction of \brog\ is known, the ratio with that of \bkg\
is proportional to $\vtdvts^2$ \cite{Ali:2004hn,Ali:2001ez}
\begin{equation}
\begin{array}{rcl}
    \frac{\mathcal{B}(\brg)}{\mathcal{B}(\bkg)} &=& S_{\rho}
    \left| \frac{V_{td}}{V_{ts}} \right|^2
    \frac{\left( 1 - m_{\rho}^2 / M^2 \right)^3}{\left( 1 - m_{K^*}^2 / M^2 \right)^3}
    \zeta ^2
    \vspace{0.2cm} \\
    & & \cdot \left[ 1 + \Delta R(\rho / K^{*}) \right] \ .
\label{eq::rhogamma:BFRatio:VtdVts}
\end{array}
\end{equation}
\noindent Here, $\zeta = \xi_{\bot}^{\rho}(0) /
\xi_{\bot}^{K^{*}}(0)$ is the ratio of form factors computed in
Heavy Quark Effective Theory (HQET), $S_{\rho} = 1 (1/2)$ are
isospin weights for the charged (neutral) $\rho$ meson and $\Delta
R(\rho / K^{*})$ is a dynamical function calculated for example in
\cite{Ali:2001ez} which accounts for different dynamics in the three
decays like vertex, hard-spectator, weak annihilation or exchange
contributions.

A recent calculation based on Light-Cone Sum Rules determines
$\xi_{\rho} = 1.17 \pm 0.09$ and $\xi_{\omega} = 1.30 \pm 0.10$
\cite{Ball:2006nr}. $\Delta R(\rho / K^{*})$ is usually computed to
be in the neighborhood of 0.1 $\pm$ 0.1. A new calculation that also
accounts for the additional $W$ annihilation diagram of the \brpg\
decay is also available \cite{Ball:2006eu}.

The same ratio of CKM matrix elements is also measured by a
physically very different process, $B$ mixing. The frequency of the
oscillation of a $B$ meson turning into a $\bar{B}$ meson is related
to the mass difference of the light and heavy $B$ meson eigenstates
$\Delta m$. The ratio of these mass differences between the
$B_{d}^{0}$ and the \bs\ is proportional to $\vtdvts^{2}$. The $B$
factories have measured $\Delta m_{d}$ very precisely
\cite{Yao:2006px} and $\Delta m_{s}$ has recently been measured by
the CDF collaboration \cite{Abulencia:2006mq}.

The first observation of \brog\ decays was reported by Belle
regarding the \brzg\ channel and the combined \brog\ channel
\cite{Abe:2005rj}. \babar\ also reported recently the observation of
the combined \brog\ and \brg\ modes and also evidence for \brzg\ and
first evidence for \brpg. This result is based on a dataset of 347
million \BB\ events~\cite{Aubert:2006pu}.

The high-energy photon is enforced to not
originate from a $\pi^{0}$ or $\eta$ decay by means of a
two-dimensional likelihood comprised of the energy of any other
photon found in the event and the two-photon invariant mass.
Continuum background is rejected with a
high-dimensional neural network, optimized for each mode
individually. These neural networks combine event shape information,
separation of the decay vertices of the two $B$ mesons in the event
along the $z$ axis, decay angular information, information related
to tagging of the flavor of the other $B$ meson in the event, and
other quantities. The signal extraction is performed with an
unbinned maximum likelihood fit in four (five) dimension for the
individual \brg\ (\bog) channels, where the dimensions used are
\mes, \de, a transformation of the neural network output, and the
cosine of the helicity angle of the vector meson (and the sine of
the second independent angle in the three body $\omega \to \pi^+
\pi^- \pi^0$ decay). For the extraction of the combined results a
simultaneous fit is used with the additional constraint on the decay
widths $\Gamma_{B\to\rho^+\gamma} = 2\Gamma_{B\to\rho^0\gamma} =
2\Gamma_{B\to\omega \gamma}$.

The resulting branching fractions are listed in
Table~\ref{tab::rhogamma:results}. While writing these conference
proceedings, the \belle\ collaboration presented an updated
preliminary analysis of these decays at the Lepton-Photon 2007
conference performed with a data sample of 657 million \BB\ events
\cite{Belle:2007lp}. Evidence is found for the two $\rho \gamma$
modes and both combined modes are observed, but the \bog\ mode still
eludes detection. These preliminary branching fractions are also
listed in Table~\ref{tab::rhogamma:results}.

\begin{table}
\renewcommand{\arraystretch}{1.1}
\centering \caption{\label{tab::rhogamma:results} The branching
fraction $(\mathcal{B})$ for each \brog\ mode is shown for the
results from \babar~\cite{Aubert:2006pu} and the new preliminary
results from \belle~\cite{Belle:2007lp}, including their
significance ($\Sigma$) in standard deviations (systematics
included).}
\begin{tabular*}{\linewidth}{l|c|c}
\hline \hline
&    \multicolumn{2}{c}{$\mathcal{B}\pm\sigma_{{\rm stat}} \pm\sigma_{{\rm sys}} (10^{-6})$ ($\Sigma$)}  \\
Mode  &  \babar  &  \belle \\
\hline
$\rho^+ \gamma$                  & $\BFrp$ (3.8) & $0.86^{+0.30 +0.07}_{-0.28 -0.08}$ (3.2)  \\
$\rho^0 \gamma$                  & $\BFrz$ (4.9) & $0.76 \pm0.17 \pm0.06$ (4.9) \\
$\omega \gamma$                  & $\BFom$ (2.2) & $0.42^{+0.20}_{-0.18} \pm0.04$ (2.6)   \\
\hline
$(\rho/\omega)\gamma$ & $\BFav$ (6.4) & $1.13 \pm0.20 \pm0.11$ (5.9)   \\
$\rho\gamma $                    & $\BFavrhorho$ (6.0)& $1.19 \pm0.24 \pm0.12$ (5.5)   \\
\hline \hline
\end{tabular*}
\end{table}

Combining both measurements to a new preliminary world average and
using the measured results of the \bkg\ decay \cite{Yao:2006px},
which includes the latest \babar\ measurement \cite{Aubert:2004te},
and applying Eq.~\ref{eq::rhogamma:BFRatio:VtdVts} with input from
\cite{Ball:2006nr,Ball:2006eu} yields the results
\begin{equation}
\begin{array}{rcl}
    \vtdvts_{\rho/\omega}^{{\rm WA}} &=& \VtdVtsvalEx \\
    \vtdvts_{\rho}^{{\rm WA}} &=& \VtdVtsBabarRhopzEx
\label{eq::rhogamma:VtdVts:result}
\end{array}
\end{equation}
for the $\rho/\omega$ mode and $\rho^+/\rho^0$ mode, respectively.

\begin{figure}
\begin{center}
    \includegraphics[width=0.44\textwidth,height=!]{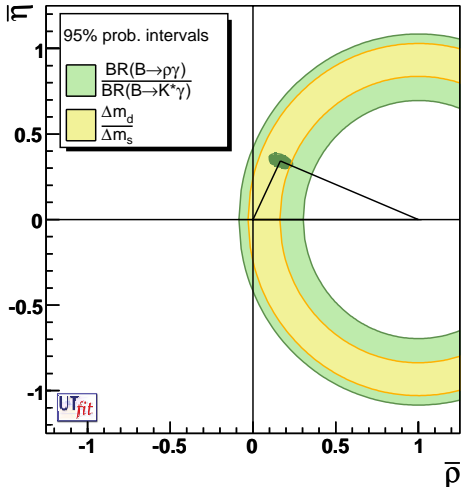}
\end{center}
\caption{Constraints on the far side of the unitarity triangle
extracted from the \brg\ measurements reported in this paper (green)
compared to $B$ oscillation measurements (yellow), computed with the
UTFit code~\cite{Ciuchini:2000de,Parodi:1999nr}.}
\label{fig::brg:ut}
\end{figure}

Both of these results are in excellent agreement with the result
extracted from $\bd/\bs$ oscillations recently reported by the CDF
collaboration \cite{Abulencia:2006mq}
\begin{equation}
    \vtdvts_{\Delta m_d/\Delta m_s}^{{\rm CDF}} = \VtdVtsCDFEx .
\label{eq::Conclusions:VtdVtsCDF}
\end{equation}
The implications of the \vtdvts\ measurement using the two $\rho$
modes only is shown in Fig.~\ref{fig::brg:ut}, compared with the CDF
result. It is a remarkable success of the Standard Model that these
two very different physics processes agree that beautifully.

\section{A new approach to \bsg}
\label{sec::bsgamma}

Besides the possibility of constraining extensions of the SM, this
decay is interesting because the high-energy photon carries
information about the $b$ quark inside the $B$ meson. If we consider
for a moment the decay of a free $b$ quark to a free $s$ quark and
the high-energy photon, then this photon has a specific energy since
this would be a two-body decay. As quarks are not free, the energy
of the photon is smeared. But still, its first moment is related to
the mass of the $b$ quark and its second moment to the Fermi motion
of the $b$ quark inside the $B$ meson. Extracting these quantities
from the high-energy photon energy spectrum feeds into the
determination of the CKM matrix element $V_{ub}$ since the extracted
parameters are universal for the $B$ meson and are needed for the
determination of this CKM matrix element from semi-leptonic $B$
decays.

The three recently published theoretical SM branching fraction
predictions at
NNLO~\cite{Misiak:2006zs,Becher:2006pu,Andersen:2006hr} show the big
interest of the physics community in this decay. These precise
calculations (up to 7\% uncertainty) allow for a very good
comparison with the measurement and thus a stringent test of the SM
can be performed.

Traditionally there are two experimental approaches to this
measurement, a fully inclusive and a sum-of-exclusive. In the fully
inclusive approach only the high-energy photon of the signal $B$
decay is reconstructed and the event is tagged as a \FourSBB\ event
by requiring the existence of a lepton ($e$ or $\mu$) coming from a
semi-leptonic decay of the other $B$ meson (the ``tag $B$'') in the
event. This method is theoretically very clean but has two
experimental drawbacks: high background and the measurement of the
high-energy photon energy spectrum in the \FourS\ rest frame instead
of the $B$ rest frame. The sum-of-exclusive approach reconstructs
the decay in as many final state modes of the $s$ quark
fragmentation as possible. Thus the photon energy is measured in the
$B$ rest frame and the background is much reduced \wrt\ the fully
inclusive method. However the theoretical interpretation is more
difficult due to the uncertainty of the missing fraction of the $s$
quark hadronization.

A new experimental approach presented by \babar\ is combining the
advantages of both analysis. The tag $B$ meson is fully
reconstructed in an all-hadronic final state, \eg\ \bdpi. Only the
photon is reconstructed from the signal side $B$ meson. Due to the
knowledge of the initial beam conditions and the full reconstruction
of the tag $B$ meson, the four momentum of the signal side $B$ meson
is known and thus, the photon energy can be measured in the $B$
meson rest frame. The fully tagged event also reduces the background
considerably. No uncertainty from the $s$ quark hadronization is
present since only the high-energy photon is reconstructed and not
the hadronization products of the signal side $s$ quark. The
drawback of this new method is the small reconstruction efficiency
of only about 0.3\%.

After the tag $B$ meson reconstruction, the signal side high-energy
photon is required to not be a product of a $\pi^0$, $\eta$, or
$\rho$ decay. The continuum background is suppressed by means of a
Fisher discriminant which is built with 12 inputs, mostly based on
event shapes. The remaining continuum background does not peak in
\mes\ and can thus be subtracted. The left over \eebb\ background is
mostly due to a $\pi^0$ or $\eta$ decay where the signal high-energy
photon is not identified as a decay product of these particles. The
signal yield is determined from an individual fit to \mes\ for each
100\mev\ photon energy bin.

The branching fraction for a minimum energy of the photon of
1.9\gev\ is determined on a \babar\ dataset with an integrated
luminosity of 210~fb$^{-1}$ as
\begin{equation}
\label{eq::bsg:bf}
    {\cal B}\left(\bsg\right)\left[E_{\gamma} > 1.9\gev\right]
    = \left( 3.66 \pm 0.85 \pm 0.59 \right) \times 10^{-4} .
\end{equation}
Extrapolating this to the theoretically preferred photon-energy cut
of 1.6\gev\ using \cite{Buchmuller:2005zv} gives
\begin{equation}
\label{eq::bsg:bfextr}
    {\cal B}\left(\bsg\right)\left[E_{\gamma} > 1.6\gev\right]
    = \left( 3.91 \pm 0.91 \pm 0.63 \right) \times 10^{-4} .
\end{equation}
The systematic uncertainties can be reduced with higher statistics.
The resulting first and and second moments for different photon
energy cuts are shown in Table~\ref{tab:bsg:moments}.

\begin{table}
\renewcommand{\arraystretch}{1.1}
\centering \caption{\label{tab:bsg:moments} Measurements of the
first and second moment of the photon energy spectrum in the \bsg\
decay, depending on the low cut on the photon energy
$E_{\gamma}^{{\rm cut}}$.}
\begin{tabular*}{\linewidth}{p{0.8cm}|l|l}
\hline \hline
$E_{\gamma}^{{\rm cut}}$  &  First moment  &  Second moment \\

$\left( \gev \right)$  &
Value    $\pm\sigma_{{\rm stat}} \pm\sigma_{{\rm sys}}$  &
Value    $\pm\sigma_{{\rm stat}} \pm\sigma_{{\rm sys}}$   \\

\hline

1.9  &  2.289    $\pm$0.058    $\pm$0.026  &
0.0334    $\pm$0.0124    $\pm$0.0065 \\

2.0  &  2.315    $\pm$0.036    $\pm$0.020  &
0.0265    $\pm$0.0057    $\pm$0.0024 \\

2.1  &  2.371    $\pm$0.025    $\pm$0.011  &
0.0142    $\pm$0.0037    $\pm$0.0013 \\

2.2  &  2.398    $\pm$0.016    $\pm$0.006  &
0.0092    $\pm$0.0015    $\pm$0.0010 \\

2.3  &  2.427    $\pm$0.010    $\pm$0.007  &
0.0059    $\pm$0.0007    $\pm$0.0003 \\

\hline \hline
\end{tabular*}
\end{table}

With this new preliminary \babar\ measurement, a third independent
approach to measure \bsg\ is explored. Since the data samples of the
three measurements are independent, they can be combined to one
result. Currently, this new method does not reach the precision of
the other two. But I expect that unlike the other two, this approach
will benefit from more statistics, especially from very large
datasets in the multi inverse attobarn era expected at super-$B$
factories.

\section{Results from the \FiveS\ run}
\label{sec::fives}

KEKB ran also at a higher CM energy equivalent to the mass of the
\FiveS\ meson. The \belle\ detector collected 23.6~fb$^{-1}$ at this
configuration, corresponding to about 2.6 million \bs\ mesons
produced. The number of produced \bs\ mesons per inverse femtobarn
is small for two reasons. First the production cross-section of the
\FiveS\ resonance is with 0.3~nb about three times smaller as that
of the \FourS\ resonance. And only about 20\% of the \FiveS\ mesons
decay into \bsss\ pairs. The majority of strange $B$ mesons are
actually created as \bss\ mesons and they decay via $\bss \to \bs
\gamma$ with a soft photon. Besides this soft photon, \bs\
reconstruction is performed as in the \FourSBB\ case. The photon is
too soft to be reliably reconstructed, so it is ignored and only
visible in a $\approx50\mev$ downward shift in \de.

\subsection{\bsphig}
\label{sec::fives::phig}

This decay mode is the brother of \bkzg, just the spectator quark is
different. The SM theory prediction for the branching fraction is
$\left( 4 \pm 1 \right) \times 10^{-5}$ \cite{Ball:2006eu}. A recent
NNLO calculation gives $\left( 4.3 \pm 1.4 \right) \times 10^{-5}$
\cite{Ali:2007sj}.

Two oppositely charged kaons are reconstructed and combined to form
the $\phi$ candidate. A cut on the invariant mass of the two kaons
is applied with a 2.5$\sigma$ window around the nominal $\phi$ mass.
The light quark continuum background is suppressed utilizing event
shape information in the form of modified Fox-Wolfram moments. The
photon is required to not originate from a $\pi^0$ or an $\eta$
decay. The final signal yield is extracted with a three-dimensional
unbinned maximum likelihood fit. The three variables entering in
this fit are \mes, \de, and the cosine of the helicity angle of the
$\phi$ meson. 18 signal events are found which yields
\begin{equation}
\label{eq::bsphig:bf}
    {\cal B}\left(\bsphig\right)
    = \left( 5.7^{+1.8+1.2}_{-1.5-1.7} \right) \times 10^{-5} .
\end{equation}
The significance of this result, including systematic uncertainties,
is 5.5$\sigma$. This is the first observation of a radiative penguin
decay of the \bs\ meson.

\subsection{\bsgg}
\label{sec::fives::gg}

This decay proceeds in the SM at leading order via a penguin fusion
diagram between the $\bar{b}$ quark and the $s$ quark. It is like
the Feynman diagram of the \bsg\ decay turned by 90$^{\circ}$ with
an additional photon attached. The SM theory predictions are in the
range $\left( 0.5 - 1.0 \right) \times 10^{-6}$
\cite{Reina:1997my,Bosch:2002bv}. This low branching fraction can be
enhanced by up to one order of magnitude in several models extending
the SM, \eg\ in 4$^{th}$ quark generation models \cite{Huo:2003cj}
or in supersymmetric models with broken R-parity
\cite{Gemintern:2004bw}. No discovery of this decay has been
reported as of today.

The reconstruction is performed similar to the above case.
Especially the two photons are required to not be consistent with
the decay of a $\pi^{0}$ or $\eta$ meson. The signal is extracted
with a two-dimensional unbinned maximum likelihood fit where the two
variables are \mes\ and \de. No signal is seen and an upper limit is
set at a 90\% confidence level of
\begin{equation}
\label{eq::bsgg:bf}
    {\cal B}\left(\bsgg\right) < 8.6 \times 10^{-6}\ {\rm at}\ 90\% .
\end{equation}
This is as of today the best limit for this decay mode.

\section{Conclusions}
\label{sec::conclusions}

New measurements of \brog\ are available. The agreement with \bs\
oscillation measurements is excellent and a non-trivial test of the
SM succeeds. The experimental precision will improve with larger
datasets. A new reconstruction method for the \bsg\ decay is
presented. This method will fully benefit from datasets from super
$B$ factories unlike previous reconstruction methods. The first
radiative penguin $B$ decay of the \bs\ meson, \bsphig, has been
observed by \belle\ using data collected at the \FiveS\ resonance.
All measurements presented have a large impact on SM tests and
constraints of beyond the SM physics. The rich field of radiative
penguin $B$ decays will provide more input to constraining the
physics beyond the SM.

%
\bibliographystyle{h-physrev4}
\bibliography{koeneke_susy07}
%
%
%

\end{document}
